\documentclass[12pt]{iopart}

\usepackage{graphicx}
\usepackage{dcolumn}
\usepackage{bm}
\usepackage[usenames]{color}    

\def\IJQI{ {\it Int. Journ Quant. Inf.}}
\def\OL{{\it Opt. Lett.}  }
\def\OE{{\it Opt. Exp.}  }

\def\PRL{{\it Phys. Rev. Lett.} }
\def\PRA{{\it Phys. Rev.} A }
\def\PRE{{\it Phys. Rev.} E }
\def\PRX{{\it Phys. Rev.} X }

\def\IJQI{{\it Int.Journ. Quant. Inf. }  }

\newcommand{\be}{\begin{equation}}
\newcommand{\ee}{\end{equation}}
\newcommand{\bea}{\vspace{0.25cm}\begin{eqnarray}}
\newcommand{\eea}{\end{eqnarray}}

\begin{document}

\title[]{Real applications of quantum imaging}

\author{Marco Genovese}

\address{INRIM, strada delle Cacce 91, 10135 Turin, Italy and CNISM, sezione di Torino Universit\`a, via P.Giuria 1, 10125 Turin, Italy}
\ead{m.genovese@inrim.it}
\vspace{10pt}
\begin{indented}
\item[]
\end{indented}

\begin{abstract}
In the last years the possibility of creating and manipulating quantum states of light has paved the way to the development of new technologies exploiting peculiar properties of quantum states, as quantum information, quantum metrology \& sensing, quantum imaging ...

In particular Quantum Imaging addresses the possibility of overcoming limits of classical optics by using quantum resources as entanglement or sub-poissonian statistics.
Albeit quantum imaging is a more recent field than other quantum technologies, e.g. quantum information, it is now substantially mature for application.
Several different protocols have been proposed, some of them only theoretically, others with an experimental implementation and a few of them pointing to a clear application. Here we present a few of the most mature protocols ranging from ghost imaging to sub shot noise imaging and sub Rayleigh imaging.

\end{abstract}

%
%
%
%
%

\section{Introduction}

In the last decade the possibility of manipulating single quantum states, and in particular photons, fostered the realization of technologies exploiting the peculiar properties of these systems (and in particular quantum correlations \cite{prep}), collectively dubbed quantum technologies \cite{4,5,6,7,8,9,10,11,prl2,ml,2,sss,gio:11,s}.

Among quantum technologies, quantum imaging \cite{s,lugr} addresses the possibility of beating the limits of classical imaging by exploiting the peculiar properties of quantum optical states \cite{Kolobov,Treps:03,boy:08,bri:10,gi}.

In this topical review we would like to present, without any pretension to be exhaustive, some of the most interesting examples of quantum imaging  protocols  that can find application in a near future. We will introduce their main features, discuss the quantum resources really needed for their implementation and, finally, describe the present state of the art of their development.

Section 2 will be devoted to ghost imaging. After introducing this technique, we will clarify what resources are really needed for this protocol and what it is achievable with classical resources. Then, a few practical applications will be presented.
In section three, we will introduce sub shot noise imaging. Here we will also discuss the resources needed for this scheme.
In section four, quantum illumination will be examined, comparing the original theoretical idea with experimental realizations.
In section five, we will review a few schemes for sub Rayleigh imaging based on quantum light.
Finally, we will hint at a few more possible developments of quantum imaging, drawing some conclusion on the future of this quantum technology.

\section{The ghost imaging}

Among quantum imaging techniques, ghost imaging \cite{gh1} (GI) was one of the first to be proposed and attracted a large interest.
This technique exploits intensity correlation fluctuations  for imaging an object
crossed by  a beam that is revealed by a detector without any spatial resolution (bucket
detector).  The image of the object is retrieved when the bucket
detector signal is correlated with the signal of a spatial
resolving detector measuring a light beam (reference beam) whose noise (the speckles pattern \cite{sp,21,22}) is spatially
correlated to this one.

This result may seem amazing sice the image is reconstructed by the spatial information gathered by the detector that did not actually interacted with the object. Nevertheless, even at an intuitive level this result can be understood from the fact that one is correlating in several different frames the fluctuations of the speckle structure of the reference beam with the effect of absorption (reflection) pattern of the imaged object on the correlated beam speckles structure.

More in detail, for realising GI, a spatially incoherent
beam is addressed to an object and then collected by a bucket
detector without any spatial resolution. The correlated beam,
not interacting with the object,  is sent to a
spatially resolving detector (an array of pixels); $\mathcal{K}$ frames are collected.
The image of the object is retrieved by measuring a function $S(x_{j})$ ($x_{j}$
being the position of the pixel $j$ in the reference region):

$$
S(x_{j})=f(E[{N}_{1}],E[N_{2}(x_{j})],E[{N}_{1}N_{2}(x)],E[{N}_{1}^{2}],E[N_{2}^{2}(x_{j})],...),
$$
where $E[{N}_{1}^{p}N_{2}^{q}(x_{j})],\,(p,q\geq0)$ is the correlation function of ${N}_{1}$, the total number of photons collected at
the bucket detector,  and of $N_{2}(x_{j})$, the photon number at the j-th pixel of the
reference arm; $E[X]$ being the mathematical average over the set of $\mathcal{K}$  frames.

Substantially, in most of experimental works three main functions $S$ have been used:

i) the Glauber intensity correlation function
\be
S(x)=G^{(2)}(x)\equiv E[{N}_{1}N_{2}(x)]\ee

ii) the normalized intensity CF
 \be S(x)=g^{(2)}(x)\equiv
G^{(2)}/(E[{N}_{1}]E[{N}_{2}(x)]) \ee

iii) the covariance, or CF of intensity fluctuations,

\be S(x)\equiv
Cov(x)=E[({N}_{1}-E[{N}_{1}])(N_{2}(x)-E[N_{2}(x)])]= \nonumber \\  E[{N}_{1}N_{2}(x)]-E[{N}_{1}]
E[N_{2}(x)] \ee

In order to discuss the performance of different ways for realising the GI, the most significant figures of merit are the resolution and the signal-to-noise
ratio (SNR) \cite{Agafonov,ESpra2009,Basano,Chan2010}
\begin{equation}
\label{SNRdef}
SNR_{S}\equiv\frac{|\left\langle S_{in}-S_{out}\right\rangle|}
{\sqrt{\left\langle\delta^{2}(S_{in}-S_{out})\right\rangle}},
\end{equation}
where $S_{in}$ and $S_{out}$ are the intensity values of the
reconstructed ghost image, when $x_{j}$ is either inside  or outside  the object profile,
\textbf{$ \delta S\equiv S-\langle S\rangle$}  is the fluctuation. The $SNR_{S}$ quantifies how well the image of the object is
distinguishable from the background. On the other hand, the resolution is determined by the number of speckles
contained in the image and quantifies the number of elementary details of the object that can be distinguished.

In view of practical application, several studies addressed a clear theoretical description
\cite{ESpra2009,exp6,per,m,shp,ipm,cai} and eventual improvements of this protocol \cite{imp,g3,Chen,boy,ipsc}, as the use of high order correlation functions or   differential schemes.

In particular, a systematic analysis, both theoretical and experimental, was presented \cite{gi},  demonstrating, among other results, that the cases (ii) and (iii) perform very similarly in terms of  $SNR_{S}$, while the case (i) performs much worse.

In the original proposal \cite{gh1} and in the earliest experimental
demonstrations of this technique \cite{gh2} nonclassical
states of light, known as twin beams, were used. These states, produced by a non-linear optical phenomenon dubbed
parametric down conversion \cite{av}, PDC, (or eventually in other non-linear optical phenomena as 4-wave mixing) are correlated in photon number and are of the form
\be |twin \rangle = \Sigma_n C_n | n \rangle | n \rangle  \label{twin}
\ee
where $| n \rangle$ is the n photons Fock state. They present a thermal statistics at single mode level and a speckle (each speckle corresponding to a spatial mode) structure at spatial multimode level.

Nevertheless, it was later shown, both
theoretically and experimentally, that GI can be realised
also with beam-split thermal light
\cite{g,gh31,gh32,gh33,gh34,gh35,gh36}, eventually even sunlight \cite{sun}, although with a smaller visibility.
Indeed by considering that the field component obeys
\be E^{+}_i ({\bf r } ,t) =  \int d^2 r'  \, h({\bf r},{\bf r'})  \int {d^3 k e^{-i {\bf k r'}} a({\bf k})} \label{E}\ee
where $a({\bf k})$ is the field annihilation operator for the optical mode ${ \bf k}$ and $h({\bf r},{\bf r'}) $ the response function of the optical path,
the correlation function of intensity fluctuations can be derived for the thermal case:
\be
Cov_{TH}({\bf x}) \propto \int_{bucket \, area} d{\bf y} | \int d{\bf x'} d{\bf y'}     h_1^*({\bf x},{\bf x'})  h_2({\bf y},{\bf y'}) \langle a^{\dag}(\textbf{x'}) a({\bf y'}) \rangle |^2
\ee
where $h_1$ and $h_2$ are the response functions on the path directed to the spatial resolving and the bucket detector respectively, while $ a({\bf x}) =   \int {d^3 k e^{-i {\bf k x}}  a({\bf k})}$.

On the other hand for twin beams:
\be
Cov_{TB}({\bf x}) \propto \int_{bucket \, area} d{\bf y} | \int d{\bf x'} d{\bf y'}     h_1({\bf x},{\bf x'})  h_2({\bf y},{\bf y'}) \langle a^{\dag}(\textbf{x'}) a({\bf y'}) \rangle |^2
\ee
A part some numerical factor and the presence of $h_1^*$ instead of $h_1$ the two CF of intensity fluctuations are analogous. In particular \cite{g}, when $T({ x})$ is the transmission function of the object to be imaged followed by a lens at a focal distance $f$ from the object and from the detection plane, $h_2({\bf y},{\bf y'}) \propto \exp(- {2 \pi i \over \lambda f} {\bf y} \cdot {\bf y'}) T({\bf y'} )$. When on the other path only a lens  is present at a distance 2f from both the source and the detection plane, $h_1({\bf x},{\bf x'} ) = \delta(  {\bf x} + {\bf x'}) \exp(- {i \pi \over \lambda f} |{\bf x'}|^2) $; by assuming that smallest scale of variation of the object spatial distribution is larger than the coherence size (speckle size), since $\langle a^{\dag}(\textbf{x'}) a({-\bf y}) \rangle $ is non-zero in a small region around ${\bf x'} = {\bf - y}$, one finds $Cov({\bf x}) \propto |T(-{\bf x})|^2$ for both cases.

In this sense, GI does not represent a "true" quantum imaging protocol, since it does not strictly require quantum light for being realised. Anyway, it must be mentioned that split thermal light has a non-zero discord, where non-zero discord is a weaker form of quantum correlation respect to entanglement \footnote{For a bipartite (A,B) quantum state $\rho$  the total amount of
correlations is defined by the quantum mutual information
\begin{equation}\label{eq:QMI}
I(\rho) = S(\rho_A) + S(\rho_B) - S(\rho),
\end{equation}
where $S(\rho) = -\Tr[\rho \log_2 \rho]$ denotes von Neumann entropy. An alternative version of the quantum
mutual information is
\begin{equation}\label{eq:CCor}
J_A = S(\rho_B) - \min \sum_k p_k S(\rho_{B \vert k}).
\end{equation}
where $\rho_{B \vert k}={Tr}_A[\Pi_k\rho\Pi_k]/
{Tr}[\Pi_k\rho\Pi_k]]$ is the conditional state of system
$B$ after obtaining outcome $k$ on $A$, $\left\{\Pi_k\right\}$ are
projective measurements on $A$,  $p_j={Tr}[\Pi_k\rho\Pi_k]]$
being the probability of obtaining the outcome $k$.
While these two definitions are equivalent in classical information,
the difference between them in quantum case defines quantum discord
\begin{equation}\label{eq:Disc}
D_A(\rho) = \mathcal{I}(\rho) - \mathcal{J}_A(\rho).
\end{equation} }, eventually to be seen as the resource needed for GI \cite{ghga}.
 However, the fact that GI does not really strictly needs quantum resources is even more evident in the so-called computational ghost imaging \cite{cgi} (for latest developments see \cite{cgin}), where discord does not play a role \cite{shsc}. In this scheme a single light beam is used by modulating its pattern. The measured correlation is between the light detected by the bucket detector and the modulation pattern.  This scheme also highlights the connection between GI and compressive sensing techniques.

In summary GI was proposed as one of the first quantum imaging examples, however its realisation does not require photon number entangled light, even if this presents some advantage, i.e. a larger visibility.

Without sticking to the discussion about "quantumness" of GI,  one can consider the practical relevance of this technique.
GI can be useful in the presence of phase distortions, where spatial information can be retrieved anyway \cite{rM}.
This means that GI is mainly of interest  in presence of a diffusive medium, a condition that appears in several significant cases, ranging from imaging in open air conditions in presence of fog to imaging of biological samples where tissues represent the diffusive medium.

In the last years, progresses toward real application of the method have been realised. In particular, several works addressed the implementation of ghost imaging in turbid media \cite{tur};
in \cite{al} was presented an application to magneto-optical Faraday microscopy, achieving an imaging of magnetic domains in garnet.

Very last experimental developments concerned GI at long distance by using optical fibers \cite{gil1}, GI ladar \cite{gil2}, GI at very low illumination level \cite{gil3,gil4}, storing and retrieval in a quantum memory \cite{yogi}...

\begin{figure}[!t]
\begin{center}
\includegraphics[width=15 cm,  bb=0 0 1200 800]{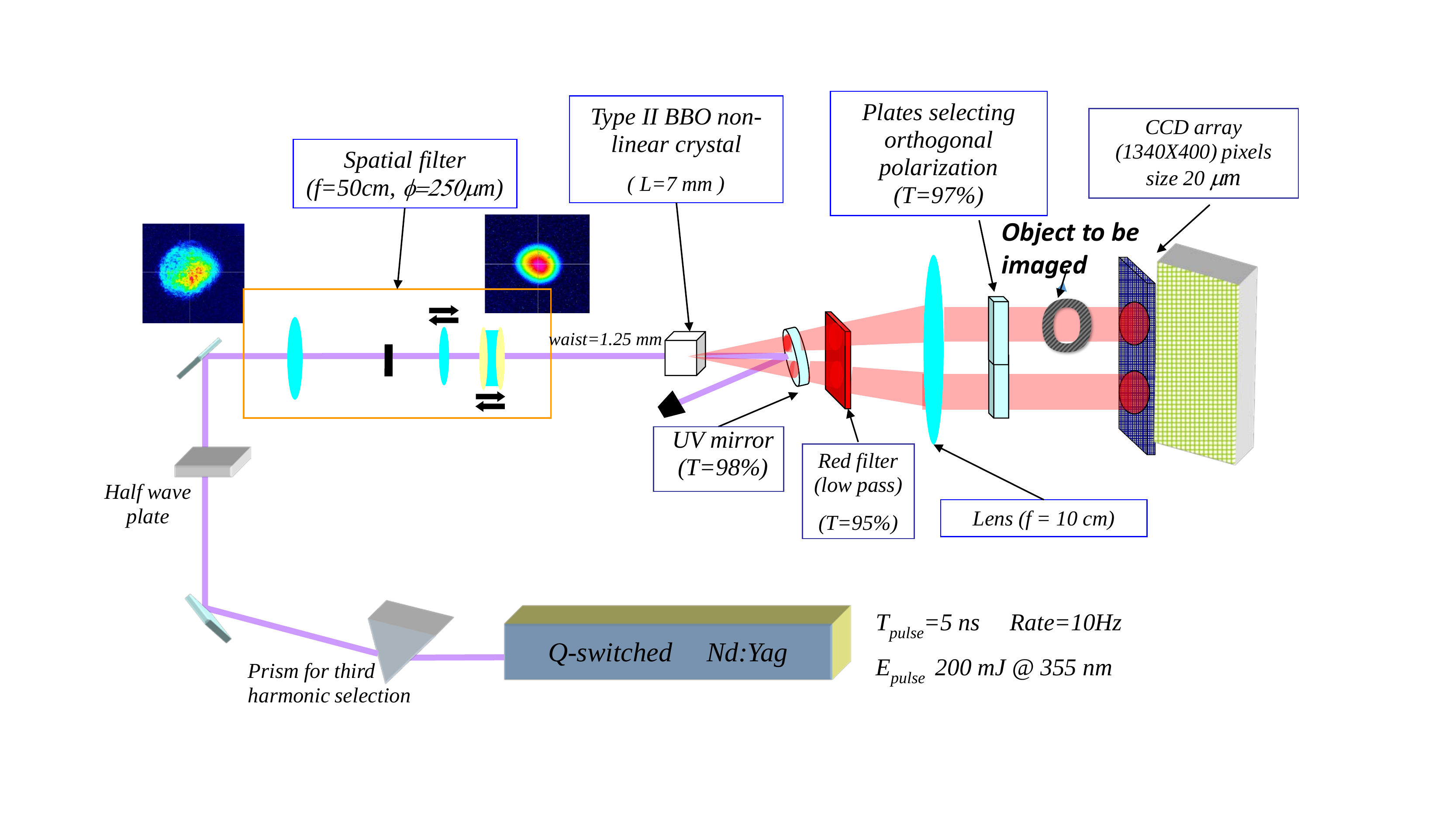}
\caption{Experimental set up of \cite{ssnexp}. A 355 nm laser beam, after spatial filtering, pumps a type II BBO crystal producing twin beams. After eliminating UV beam, one correlated beam crosses a weak absorbing object  and it is then addressed to  a CCD camera. The other beam is directly sent to another area of CCD camera. }\label{fig1}
\end{center}
\end{figure}

\section{The sub shot noise quantum imaging}

One fundamental limit of measurement is the so called shot noise, deriving from the quantisation of carriers as photons in light beams and electrons in electric current.
For classical light the limit is reached by the Poissonian statistics of coherent states, i.e. the  standard deviation of shot noise is equal to the square root of the average number of photons N.
For other classical states of light (e.g. thermal states) the variance is larger (super Poissonian statistics).

For what concerns imaging, this means that, due this unavoidable noise deriving from the fluctuations of the photon number, it is impossible to image a very weakly absorbing object under a certain low illumination level, for this is lost in the noise.

A very interesting scheme for beating this limit was proposed  \cite{lug}, suggesting to use photon number correlations between two light beams for measuring the noise in a reference beam and subtracting it from the imaging beam. In particular, as we mentioned in the previous section, twin beams (\ref{twin}) have each a thermal statistics, but the fluctuations in photon number are exactly the same. Thus, the noise, in an ideal situation, can be completely eliminated  by measuring it in one beam and subtracting to the other. Nevertheless, in a realistic situation the effectiveness of this method depends on the detection probability, since quantum correlations are spoiled by losses.

The effectiveness of reaching sub poissonian regime is quantified by the noise reduction factor (NRF)
\be \sigma\equiv \langle\delta^{2}(N_{1}-N_{2})\rangle/\langle N_{1}+N_{2}\rangle\approx 1-\eta
\ee
where $N_i$ is the photon number in beam i=1,2 and $\delta N =N-\langle N\rangle$ ($\eta$ being the total detection efficiency \footnote{Incidentally this linear dependence can be exploited for ccd calibration \cite{cal}.}).

The subpoissonian (non-classical) regime, $\sigma < 1$, has been reached in several works \cite{s1,s2,s3,s4,s5,s6P,s6,s7}, but in conditions that did not allow for imaging (being limited to a single spatial beam or requiring background subtraction).

A multimode subpoissonian regime without any background subtraction was finally reached \cite{ssn}, paving the way to the first experimental demonstration of this scheme \cite{ssnexp,ssnpra}. In this last work (see fig.1) two twin  beams, produced by generating with a 355 nm laser beam (with 5 ns pulses) type II (opposite polarisations) PDC, were addressed (by a lens in a f-f configuration) to a high quantum efficiency (about 80\% at 710 nm) CCD camera synchronized with with the pump. The multi-thermal statistics of the source (with thousands of temporal modes \cite{bram}) was quasi Poissonian with excess noise close to  zero. By tayloring the mode structure \cite{21,22} a NRF $\sigma = 0.452 (0.005)$ was achieved, allowing to beat any classical protocol and retrieving an image of a thin atomic deposition on glass with an improvement of the signal to noise ratio larger than 30\% compared with the best classical imaging scheme and more than 70\% better than the differential classical scheme (split thermal light beam).

Later, the advantage of this quantum protocol was enhanced reaching a NRF=0.29 \cite{spie} and the scheme was applied to microscopic regime \cite{rb16}.

Successively the method found further developments, for example a progress toward real time sub shot noise quantum imaging was achieved \cite{rqi}.

Also in this case it is interesting trying to understand which resources are really needed for realising this protocol. Evidently, entanglement is not really needed since what is required is a strict correlation in photon number fluctuations, that would be achievable with photon number Fock states if these could be efficiently generated. Anyway, in this case the need of quantum states looks to be unambiguous.

Finally, in \cite{on13} the shot noise limit was beaten for differential interference contrast microscope, a system where horizontal and vertical polarization of light are directed to the sample through different optical paths, experiencing different phase shifts that are detected as a polarization rotation at the output. This scheme is limited by the shot noise.  In  \cite{on13} a signal to noise ration  $1.35 \pm 0.12$ times better than the Shot Noise was reached by using a NOON state $|NOON \rangle = (|0 N \rangle + |N 0 \rangle) / \sqrt{2} $ with N=2 in a confocal configuration. This proof of principle is very interesting, but the possibility of building NOON states with $ N$ significantly bigger than 2  is still beyond present possibilities, strongly limiting practical use of techniques based on these states.

\section{The quantum illumination}

\begin{figure}[!t]
\begin{center}
\includegraphics[width=15 cm,  bb=0 0 1200 800]{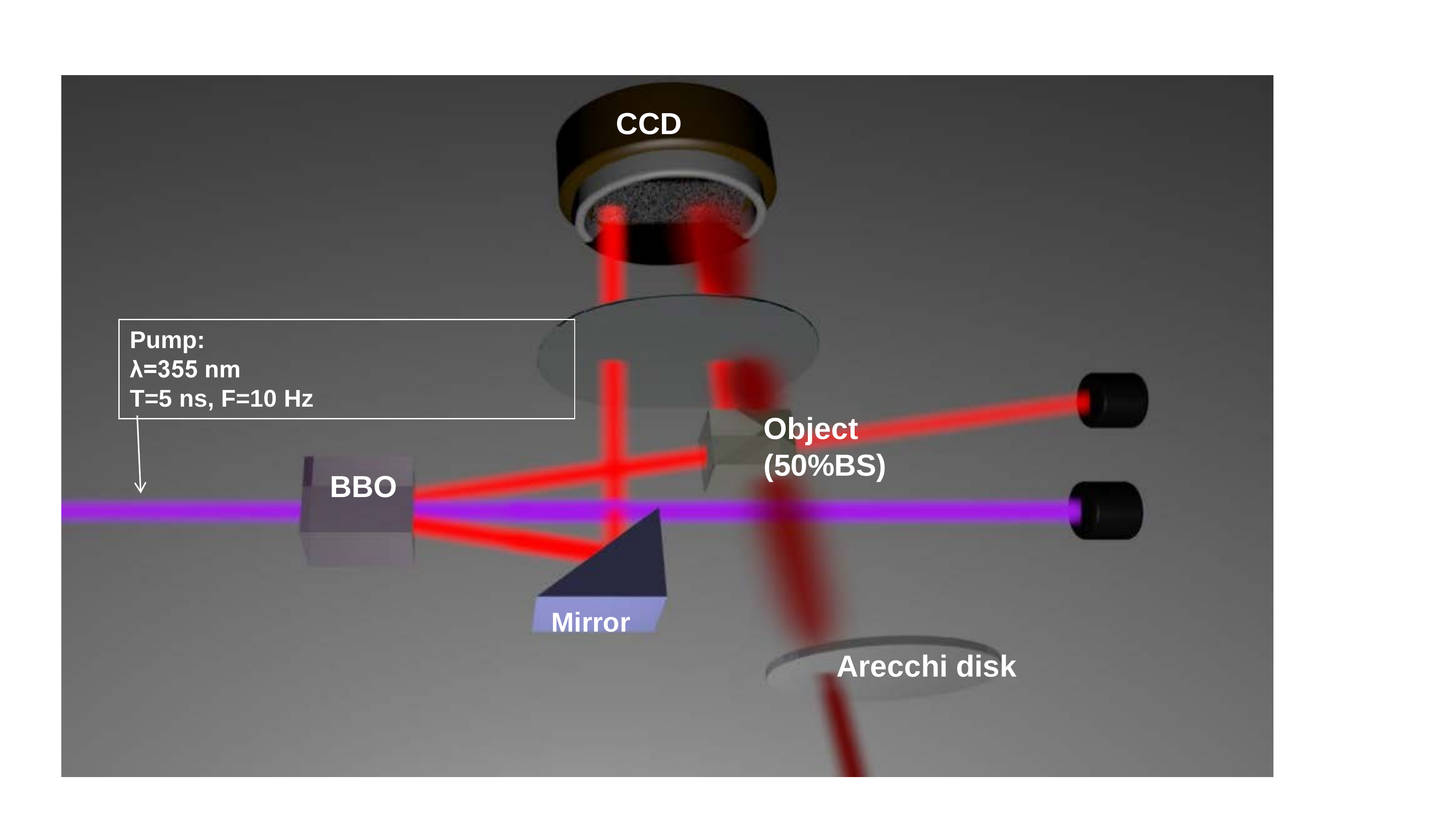}
\caption{Experimental set up of the first realisation of quantum illumination \cite{l}. A laser beam at 355 nm pumps a type II BBO crystal producing twin beams. After eliminating UV beam, the return beam is addressed to the target and combined with a thermal background, generated through an Arecchi's disk, before being addressed to a CCD camera. The other beam is directly sent to another area of CCD camera. }\label{fig2}
\end{center}
\end{figure}

Quantum illumination (QI) was introduced in  \cite{llo:08}, expanding some previous theoretical idea \cite{da}, as a method for detecting and imaging objects in the presence of high losses and background by exploiting quantum states of light.
This seminal work compared the theoretical limit for detecting an object by using single photon states in mode k, $| k \rangle$, with entangled single photons of the form $(1 / \sqrt{d} ) | \Sigma_k  | k \rangle | k \rangle $, where one mode (return mode) was addressed to the object and the other (idler mode) used as a reference.  An evident advantage of the quantum protocol was demonstrated even in presence of strong losses and noise: a result of the utmost interest since typically quantum protocols rapidly loose their advantage in this situation.
Thus, this paper not only presented a new interesting possibility of overcoming classical limits, but also demonstrated for the first time that in certain situations quantum protocols can be robust against noise and losses, a fact not at all evident considering the extreme delicacy of previous quantum protocols.

Nevertheless, no specific detection scheme was presented and from this seminal work was not possible to directly plan an experiment.
 Later a progress in this sense was realised \cite{tan:08}, considering a producible quantum optical field,  a twin beam state (\ref{twin}), jointly with an optimum joint measurement and comparing this scheme with the optimum for every classical protocol. A 6 dB advantage of the quantum one was demonstrated.

Further progresses toward a realistic implementation were presented  \cite{sha:09,ms,gua:09}. In particular,  two possible detection schemes were suggested  \cite{gua:09}. In the so called OPA receiver the return mode and the idler mode are inputs of a Optical Parametric Amplifier and the total number of photons is counted at one output port. In the Phase-Conjugate Receiver return and idler modes are inputs of a balanced difference detector.

Albeit interesting, these two detection methods rely on lossless photon counting, a detection method difficult to be realised in experiments.

A different approach was assumed in \cite{l,l1}, were the first experimental realisation of quantum illumination was realised, pointing to a simple and efficient experimental setup
showing that second order correlation measurements on return and idler twin beams already suffices in guaranteeing strong advantages to the quantum protocol.

In a little more detail (fig.2), type II Parametric Down Conversion light (with correlated photon pairs of orthogonal polarisations) was generated by means of a BBO non-linear crystal pumped with the third harmonic (355 nm) of a Q-switched Nd-Yag laser, with 5 ns pulses. The correlated emissions were then addressed to a high quantum efficiency (about 80\% at 710 nm) CCD camera.
 On one of the two paths  the target object  was posed. It consisted of a beam splitter  combining the PDC light with a thermal background produced by addressing a laser beam to an Arecchi's rotating  ground glass \cite{Ar}.
In a  second configuration, classical, the twin beam was substituted by beam split thermal light.

The exponential advantage of QI respect to the classical method is shown in Fig.3, where the error probabilities for detecting the target are reported and compared with the experimental data of \cite{l}.

\begin{figure}[tbp] \begin{center}
 \includegraphics[width=0.9\textwidth]{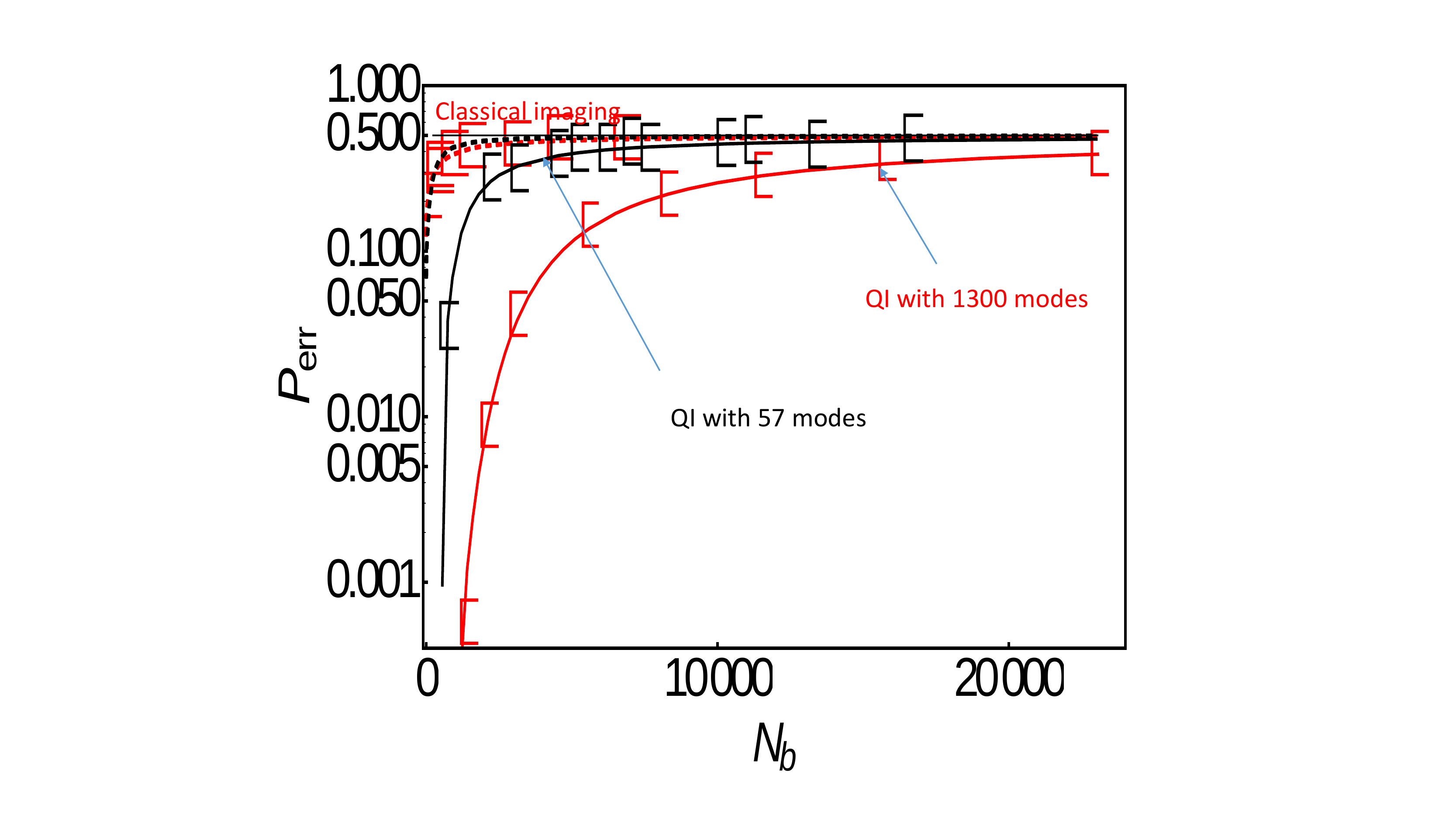}

\caption{Results of quantum illumination protocol \cite{l}: the error probability $P_{\rm err}$ of the target detection
  versus the total number of photons of the thermal background $N_{b}$
  evaluated with 10 frames. The squares are the data  \cite{l} compared with the theoretical predictions for clasiccal imaging and QI.} \label{Perr} \end{center}
\end{figure}

Which resource is really needed for this specific protocol was then discussed in a subsequent paper \cite{ad}, where it was demonstrated as the improvement in the signal-to-noise ratio achieved by the quantum sources over the classically correlated thermal ones is closely related to the respective ratio of mutual informations of return and idler beams, ${MI} (r:i)=S_2(\rho_r)+S_2(\rho_i)-S_2(\rho_{ri})\,,$ with, for a Gaussian state with covariance matrix $\boldsymbol{\sigma}$,  $S_2(\rho) = \frac12 \ln (\det \boldsymbol{\sigma})$.

A further progress was then realised  \cite{s}. In this work, the detection, albeit sub-optimal, better exploited the photon number entanglement of twin beams (produced by pumping with a cw 780 nm laser beam a periodically poled lithium-niobate crystal) combining the return and idler modes in a OPA. The method was applied to secure quantum communication.

Finally, recently a theoretical idea for an extension to microwave region was presented \cite{mB} and further theoretical ideas (as the use of photon subtracted photons) were discussed \cite{qiluut}.

\section{Sub Rayleigh Quantum Imaging}

The wave nature of light limits the resolution achievable with a microscope, as studied by Abbe and Rayleigh. The importance of high resolution in biology, material science etc. prompted the search for methods beating  Rayleigh limit (for a microscope $0.61 \lambda / NA$, NA being the numerical aperture). Significant results have been reached with classical light \cite{nob}, as stimulated emission depletion (STED) and
ground state depletion (GSD). Nevertheless, these schemes are characterized by rather specific experimental requirements (dual
laser excitation system, availability of luminescence quenching mechanisms by stimulated emission, nontrivial shaping of the quenching beam, high power); furthermore, they are not suited for applications where the fluorescence is not optically induced.

A possibility to overcome these limits or to improve these methods is offered by the correlation properties of quantum light.
As discussed in \cite{po}, the field component obeys
\be E^{+}_i ({\bf r_i } ,t) = \int {d^3 k e^{-i kt/c} [ \int d^2 r A({\bf r})  h({\bf r_i},{\bf r}) ] a({\bf k})} \label{E}\ee
where $a({\bf k})$ is the field annihilation operator for the optical mode ${ \bf k}$, $A({\bf r})$ the object aperture function and $h({\bf r_i},{\bf r}) $ the point spread function of the imaging apparatus. When considering incoherent imaging, $P^{inc}({ \bf r}) \propto \int d^2 r |A({ \bf r})  h({\bf r_i},{\bf r})|^2$, or coherent imaging,  $P^{c}({\bf r}) \propto |\int d^2 r A({\bf r}) h({\bf r_i},{\bf r})|^2$, the convolution integrals produce blurred images whose amount can be gauged through the Rayleigh diffraction bound. The idea behind sub Rayleigh imaging is beating this limit by using appropriate light sources and by replacing intensity measurements with N-fold coincidence detection strategies.
In particular one can envisage \cite{po} to achieve a resolution enhancement at the standard quantum limit ($\propto 1 / \sqrt{N}$, being N the number of photons) or at the Heisenberg bound \cite{qm} ($\propto 1 / N$) according to the procedure.

A first idea is to exploit the effective wave length of N photons systems, as biphotons produced in Parametric Down Conversion (i.e. the twin beams (\ref{twin}) in the low gain regime when only the two photons component is significantly different form zero). When measuring the 2nd order correlation function the interference presents a modulation corresponding to a wave length ($\lambda / 2$) reduced of a factor 2 respect to the one, $\lambda$, of the used light \cite{bip}.
This allows an increased resolution in two photon detection, a result that can be extended to NOON states \cite{noon} for which one expects a correlation function \be
G^N(\textbf{r}, ...,\textbf{r}) \propto 1 + \cos N \delta(\textbf{r}) \label{GN}
\ee
with a $\lambda / N$ improvement, originating a method dubbed "quantum litography" \cite{lit,xu}. Nevertheless, NOON states suffer of a strong weakness: the probability of detecting N photons arriving at the same place decreases (as the efficiency of the scheme) exponentially with N \cite{nw}. This weakness, potentially spoiling the advantage of this strategy, can be overcome with an optical centroid measurement \cite{t}. This idea was realised \cite{b} for 2 photons NOON states and then  for 2,3 and 4 photons NOON states produced with a PDC scheme \cite{r}.

Due to the difficulty in producing light with multiphoton entanglement, as NOON states, a second strategy relies on exploiting postselection to extract the "non-classical component" from a classical state containing information about the object to be imaged. Several theoretical schemes have been proposed  \cite{po,post}; in particular  the possibility of beating Rayleigh limit by exploiting N-fold coincidence detection was highlighted \cite{po}.

 A first interesting experimental realisation has been presented \cite{POST1}, where a laser pulse is split into
two equal components, then one is shifted in phase with respect to
the other, and the two components are finally allowed to interfere on a lithographic plate that
functions by means of N-photon absorption. To achieve enhanced resolution by a factor of
M, the process is repeated M times, incrementing the relative phase of successive laser pulses  by a fixed amount. The effect of averaging M laser shots with progressively increasing phase shifts is to
average out  undesired spatially varying terms, leaving only a spatially uniform
component of the form $ \cos( 2 \pi M x \chi)$ at the desired frequency ($\chi$ being the the period $2 \lambda  sin \theta$ of the 1-photon intensity pattern created by the interference
of the two beams, with $\theta$ the angular separation between the beams). Therefore, the pattern has a resolution M-times
better than that allowed by normal interferometric lithography, analogously to quantum lithography. Finally, the N-photon-absorption recording medium
was simulated by Nth harmonic generation followed by a CCD camera. Here it is worth mentioning that a primary impediment to implementing these techniques is the difficulty of developing suitable N-photon
absorbing media, especially for N large. Nonetheless, some idea in this sense is emerging, as dopplerons \cite{d} or other \cite{o}.

In \cite{post3} was  considered a short pulse exciting a narrow transition in a material such that the excitation lifetime is much longer
than the pulse duration. If the transition is excited again by another pulse within the
excitation lifetime of the first pulse, the two excitations interfere even if the two pulses
do not, being mutually incoherent. As this interference occurs through the medium, the
relative phase  is given by the transition frequency $\nu$ and the relative delay
$t$ between the pulses. When considering a non-linear excitation of order N, the center
frequency of the exciting pulse is $\nu t / N$, i.e. a  "quantum interference" corresponding to a wavelength N times smaller. A proof of principle of the method was realised with rubidium atoms reaching a factor 2 under Rayleigh limit.

In \cite{post2} super--resolution was reached by illuminating an object with a laser and by post-selecting, in a SPAD array where the image is collected, only pixels that had counted exactly a certain photon number N. Indeed, in this way the 2N power of the object field transmission function is convolved with the Nth power of the receiver's point spread function, an Airy function, which is $\sqrt{N}$ times narrower than the conventional point spread function. A similar method was used in \cite{post4}, where a focused laser beam scanned an object together with
dynamic application of  a threshold N less than the maximum count level $N_{max}$; the experimental results demonstrated a sub-Rayleigh resolution enhancement by a factor of $\sqrt{[ln(N_{max}/N)]}$.

After these seminal papers, the possibility of sub Rayleigh imaging with "quantum" post-selected classical light was further explored in several more, interesting, papers. Always in the line of \cite{po},   pseudothermal light was used with second order correlation measurement \cite{yo}. Two-photon interference with subwavelength fringes has been observed for the first time with true thermal light in \cite{wu}. Further results in this line have been presented in \cite{th}.

In general for achieving super--resolution with 2nd order correlations with biphoton states one scans the two point detectors simultaneously in the same direction, being $g^{(2)}(x,y) \propto \cos[ k (x+y)]$, while for thermal light one simultaneously scans the two point detectors opposite directions, being $g^{(2)}(x,y) \propto \cos[ k (x-y)]$: a systematic study (both theoretical and experimental) of the best way for achieving super-resolution with second order correlation was done in Ref.\cite{sc}.

Another interesting strategy is based on the use of single photon sources.  A first idea in this sense was proposed \cite{zan07} suggesting to achieve the super-resolved modulation of Eq. \ref{GN} by exploiting N photons emitted by N atoms. Then,  it was demonstrated that by measuring the second order correlation function of two (four) single photon sources (two excited atoms) illuminating the object to be imaged the resolution improves of a factor 2 (4) \cite{zan09}. The idea to exploit independent sources for achieving the super-resolved modulation of Eq. \ref{GN} was further studied, both theoretically and experimentally \cite{zan12}, considering a configuration of N independent emitters (thermal light sources) along a chain with equal spacing and N-1 detectors placed in a semicircle at specific angles (in this case no real quantum state preparation or detection is needed).

A very interesting method was proposed in \cite{or} and extended and applied to NV colour centres in \cite{pt}.  Here the idea is to exploit antibunching of single photon sources. In synthesis, if $\mathcal{P}(x)$ is the probability of detecting a photon from a single photon emitter, by taking the $k$-th power, $[\mathcal{P}(x)]^k$, the function gets narrower increasing the resolution. However, the $k$th power of the signal contains also products terms. The method consists in eliminating these product terms by subtracting high order correlation functions:

\begin{equation}
\label{eq:k2}
\sum_{\alpha=1}^2 [\mathcal{P_\alpha}(x))]^2=\langle\hat{N}\rangle^2[1-g^{(2)})]
\end{equation}

\begin{equation}
\label{eq:k3}
\sum_{\alpha=1}^3 [\mathcal{P_\alpha}(x)]^3=\langle\hat{N}\rangle^3[1-\frac{3}{2}g^{(2)}+\frac{1}{2}g^{(3)})]
\end{equation}

\begin{equation}
\label{eq:k4}
\sum_{\alpha=1}^4 [\mathcal{P_\alpha}(x)]^4=\langle\hat{N}\rangle^4\{1-2g^{(2)}+\frac{1}{2}[g^{(2)}]^2+\frac{2}{3}g^{(3)}-\frac{1}{6}g^{(4)}\}
\end{equation}

\begin{eqnarray}
\label{eq:k5}
\sum_{\alpha=1}^5 [\mathcal{P_\alpha}(x))]^5=\langle\hat{N}\rangle^5\{1-\frac{5}{2}g^{(2)}+\frac{5}{4}[g^{(2)}]^2+\frac{5}{6}g^{(3)}+ \nonumber  \\
-\frac{5}{12}g^{(2)}g^{(3)}-\frac{5}{24}g^{(4)}+\frac{1}{24}g^{(5)}\}
\end{eqnarray}

In general, the expressions of the super-resolved images for any $k$ have the following form:

\begin{equation}
\label{eq:k8}
\sum_{\alpha=1}^n [\mathcal{P_\alpha}(x)]^k=\langle\hat{N}\rangle^k\sum_{i=1}^{i_{max}}y_i \beta_{i},
\end{equation}
with $\beta_{i}$ representing products of the form $g^{(j_1)}(0) \cdot g^{(j_2)(0)} \cdot \ldots \cdot g^{(j_l)}(0)$.

The results of  \cite{pt} are summarised in fig.4, demonstrating that the Abbe limit is beaten.

 \begin{figure}[tbp] \begin{center}
 \includegraphics[width=0.9\textwidth]{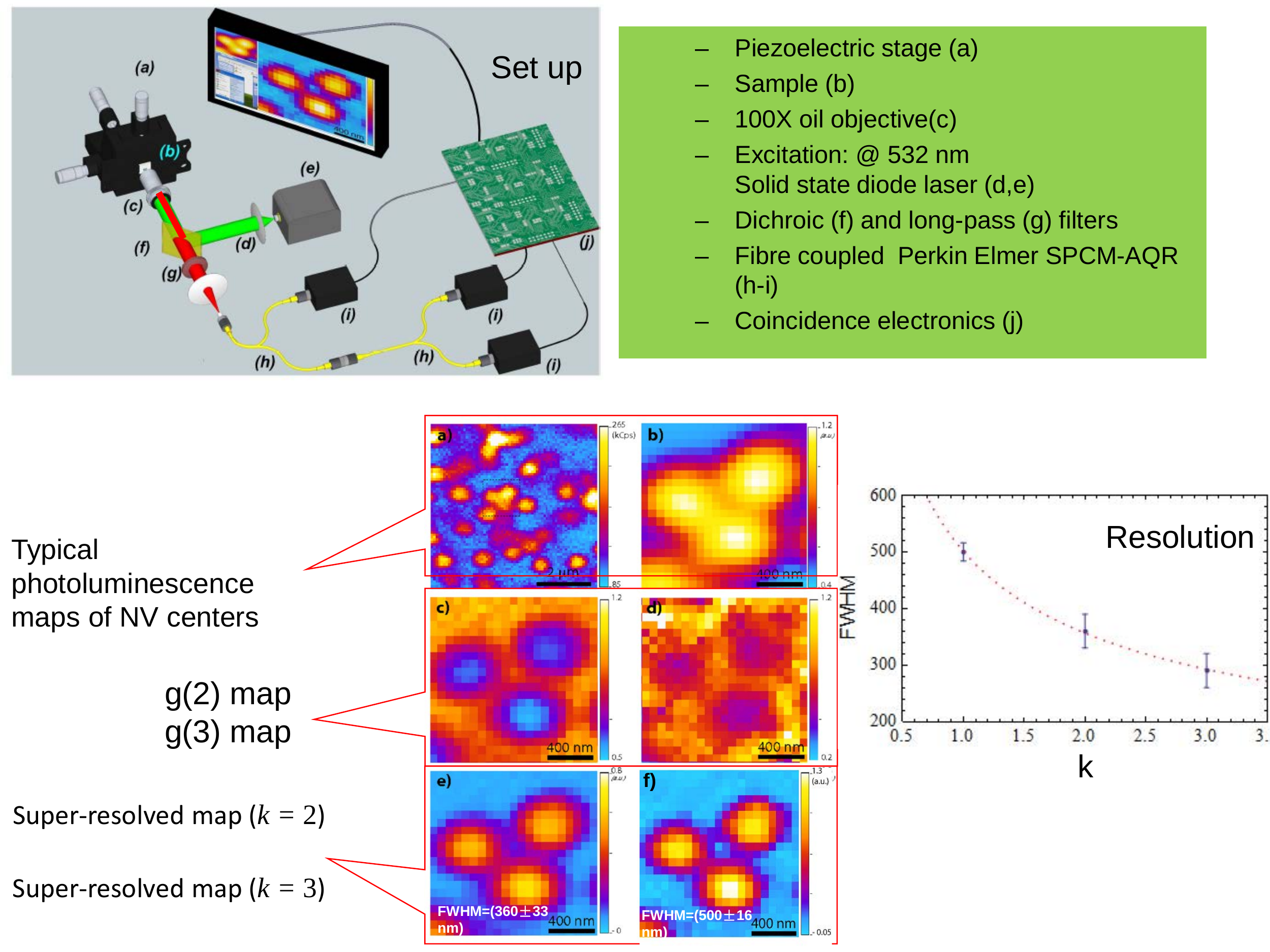}

\caption{On the top the set up of Ref. \cite{pt}: a laser scans the sample containing NV colour centers in diamond through a confocal microscope. The emission of single photon emitters is then collected and addressed to SPAD detectors. On the bottom   are shown: a) a typical map of the sample, b) the direct imaging of three unresolved NV centres, c)  and d) the $g^{(2)}$ and $g^{(3)}$ maps respectively, e) and f)  the super resolved map for k=2 and k=3 respectively. In the inset on the right the plot of the Full Width Half Maximum characterizing the resolution, improving with the order k of the correlation function.}\label{f4} \end{center}
\end{figure}
This method, due to the biocompatibility of nanodiamonds and their possible use for measuring very weak currents can find application in biology of the utmost importance \cite{v2,sr} (in fig.5  a map of single photon emission from NV centers in nanonodiamonds in neuronal cells is shown).
\begin{figure}[tbp] \begin{center}
 \includegraphics[width=0.9\textwidth]{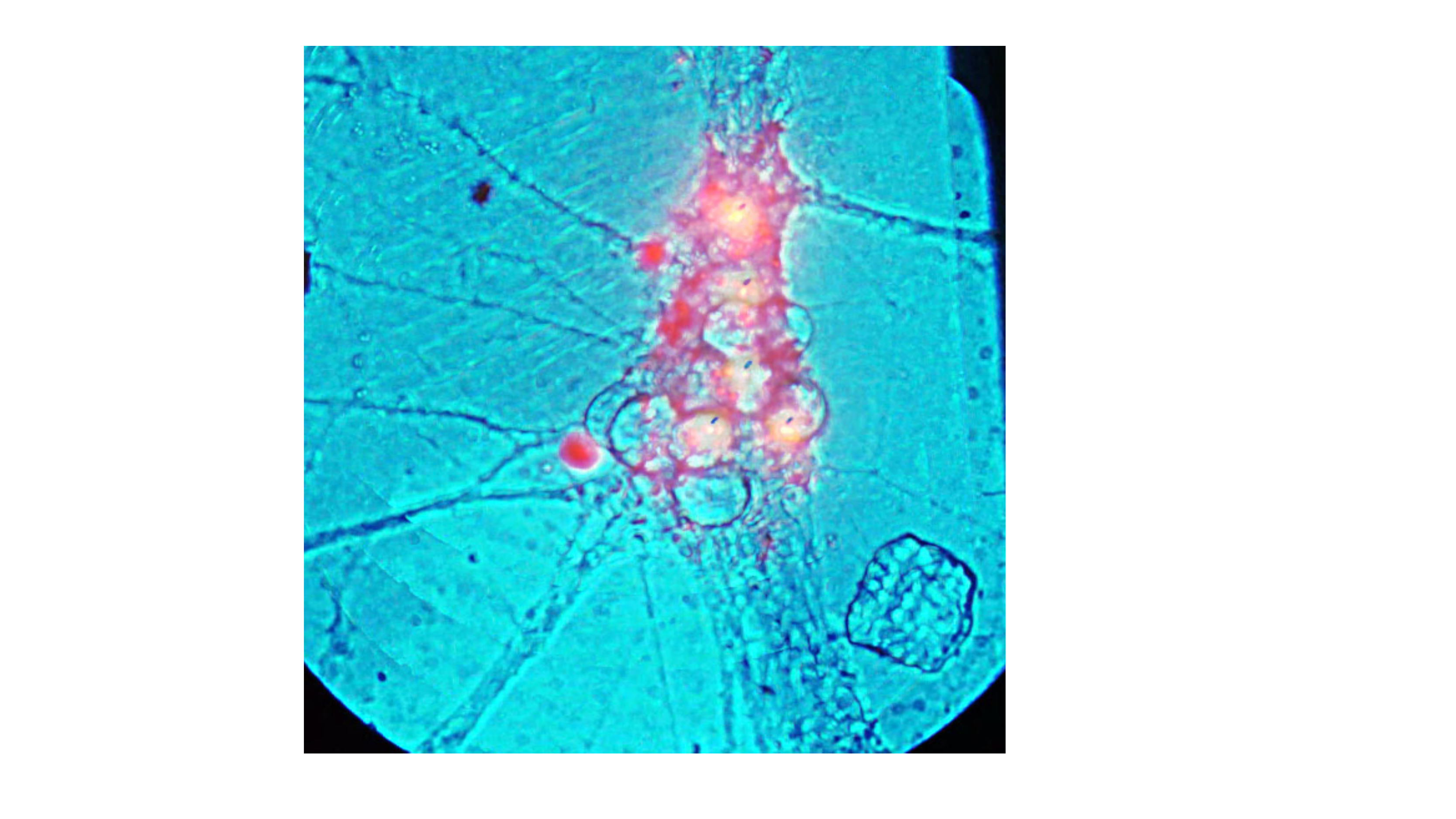}

\caption{ Emission from NV colour centres in nanodiamonds inside hippocampal neurons.}
\label{f5} \end{center}
\end{figure}

   Finally it is worth mentioning the use of squeezed light \footnote{the relation between squeezing and entanglement has been discussed, for example, in \cite{lugr}} for beating Rayleigh limit, since this allows reducing the intrinsic noise of light that limits traditional super-resolution schemes \cite{kf}. In \cite{ta13} 6 dB amplitude squeezed light was combined with a imaging field for tracking the motion of particles in a biological sample, beating of 42\% the standard quantum limit. The same scheme was then used in \cite{ta14} for realising a photonic force microscopy with a 14\% resolution enhancement respect to coherent light experiments.

\section{Perspectives and Conclusions}

In conclusion quantum imaging is emerging as one of the most interesting quantum technologies: several protocols have been proposed, some of them with significant perspectives of real application in a next future.
In this review we have described with a certain detail ghost imaging, sub shot noise imaging, quantum illumination and a few of the several proposals for beating Rayleigh diffraction limit by exploiting quantum states of light.

Being exhaustive in describing all quantum imaging proposals is rather difficult and, probably, pointless. Therefore, we limited our choice to these few examples that are somehow paradigmatic of these quantum technologies and, in our opinion, next to application. Nevertheless, before concluding, it is worth mentioning a few other protocols of quantum imaging, as entangled imaging \cite{bo08}  (i.e. the generation of two correlated images by exploiting twin beams entanglement), noiseless amplification of images \cite{mo05}, quantum imaging with undetected photons \cite{le14} an much more \cite{alf}.

Altogether we hope to have provided a panoramic view  on this new interesting quantum technologies and to have shown their potentiality for a significant development in the next future.

\subsection{Acknowledgments}
I wish to thank all my collaborators with whom I had the pleasure to collaborate in the fascinating field of quantum imaging:
A.Avella, G. Brida, A.Caprile, I. P. Degiovanni, J. Forneris,  D. Gatto Monticone, K. Katamadze, E.D. Lopaeva, A.Magni, A.Meda, E. Moreva, S.Olivares, P. Olivero, I. Ruo Berchera,N.Samantaray, P. Traina.
This work has been partially supported by EU through BRISQ2 project (EU FP7 under grant agreement No. 308803). We also acknowledge the support of the John Templeton Foundation (Grant ID 43467). The opinions expressed in this publication are those of the authors and do not necessarily reflect the views of the John Templeton Foundation.


\section{References}
\begin{thebibliography}{99}
\bibitem{prep} Genovese M, \emph{Phys. Rep.} \textbf{413}(2005) and ref.s therein.
\bibitem{4} Bouwmeester~D, Pan~J~W, Mattle~K, Eibl~M, Weinfurter~H and Zeilinger~A 1997 {\it et al.} {\it Nature} {\bf 390} 575
\bibitem{5} Ursin~R, Jennewein~T, Aspelmeyer~M, Kaltenbaek~R, Lindenthal~M, Walther~P and Zeilinger~A 2004 {\it Nature} {\bf 430} 849
\bibitem{6} Boschi~D, Branca~S, De Martini~F, Hardy~L and Popescu~S 1998  {\it Phys. Rev. Lett.} {\bf 80} 1121
\bibitem{7} O'Brien~J~L  2007 {\it Science}  {\bf 318} 1567
\bibitem{8} Yao~X~C et al 2012  {\it Nature} {\bf 482} 489
\bibitem{9} Yamamoto~T, Koashi~M, \"Ozdemir~S~K and Imoto~N 2003 {\it Nature}  {\bf 421} 343
\bibitem{10} Pan~J~W, Simon~C, Brukner~C and  Zeilinger~A 2001  {\it Nature} {\bf 410} 1067
\bibitem{11} Pan~J~W, Gasparoni~S, Ursin~R, Weihs~G and Zeilinger~A 2003  {\it Nature} {\bf 417}, 4174
\bibitem{prl2}  Ruo~Berchera~I, Degiovanni~I~P, Olivares~S and Genovese~M 2013
{\it  Phys. Rev. Lett.\/} {\bf 110} 213601
\bibitem{ml} Brida G, Castelletto S,  Degiovanni I P,  Genovese M,  Novero C, Rastello M L
2000, {\it Metrologia} {\bf 37} 629;
\bibitem{2} Brida G,  Degiovanni I P,  Florio A,  Genovese M,  Giorda P,  Meda A,
 Paris M,  Shurupov A 2010 { \it Phys Rev Lett} {\bf 104} 100501


 \bibitem{sss} Zhang Z. \etal 2013 \PRL \textbf{111} 010501
\bibitem{gio:11} Giovannetti~V, Lloyd~S and Maccone~L 2011  {\it Nat. Phot.
 } {\bf 5} 222

 \bibitem{s} Simon S \etal 2014 \IJQI \textbf{12} 1430004
\bibitem{lugr} Lugiato L A \etal 2002\emph{ J.Opt. B} \textbf{4} S176

\bibitem{Kolobov}  Kolobov~M~I 2007 \emph{Quantum Imaging} (New York: Springer)

\bibitem{Treps:03} Treps~N, Grosse~N, Bowen~W~P, Fabre~C, Bachor~H~A and  Lam~P~K 2003 {\it Science\/} {\bf 301} 5635 940

\bibitem{boy:08} Boyer~V, Marino~A~M, Pooser~R~C and Lett~P~D 2008 {\it Science\/} {\bf 321} 544

\bibitem{bri:10} Brida~G, Genovese~M and Ruo~Berchera~I 2010  {\it Nature Photonics\/} {\bf 4} 227

\bibitem{gi} Brida G et al. 2011 {\it Phys Rev. A} {\bf 83} 063807


\bibitem{gh1}Belinskii A, Klyshko D 1994 {\it Sov. Phys.JETP} \textbf{78} 259
\bibitem{sp} Berzanskis A, \etal 1999 \PRA {\bf 60}  1626
\bibitem{21} Brida G \etal 2009 \emph{Int. Journ. Quant. Inf.} \textbf{7} 139
\bibitem{22} Brida G \etal 2009  \emph{Journal of Modern Optics} \textbf{56} 201

\bibitem{Agafonov} Agafonov  I N \etal arXiv:0911.3718v2 [quant-ph]
\bibitem{ESpra2009}  Erkmen B and Shapiro J 2009 \PRA \textbf{79}, 023833

\bibitem{Basano} Basano L and Ottonello P 2007 \emph{Opt. Express }\textbf{15}, 12386
\bibitem{Chan2010}Chan K, O'Sullivan M and Boyd R 2010 \emph{Opt. Express} \textbf{18} 5562
\bibitem{exp6}Santos I,  Aguirre-Gomez J and Padua S 2008  \PRA \textbf{77},  043832

\bibitem{per}  Scarcelli G , Berardi V and Shih Y 2006 \emph{Journ. Mod. Opt.} \textbf{53 }16

\bibitem{m} Lopaeva E and Chekhova M 2010 \emph{JETP Lett.} \textbf{91}  447
\bibitem{shp}  Erkmen E and Shapiro J 2008 \PRA \textbf{77} 043809
\bibitem{ipm} Degiovanni I P  \etal 2007 \PRA \textbf{ 76} 062309
\bibitem{cai} Cai Y and Zhu S \etal 2005 \PRE \textbf{ 71} 056607

\bibitem{imp} Degiovanni I P \etal 2007 \IJQI \textbf{5} 33;  Ferri F \etal 2010  \PRL
\textbf{104} 253603.

\bibitem{g3} Brida G \etal 2011 \IJQI \textbf{9}  341 ; Bai Y and Han S 2007 \PRA 76
043828; Zhou Y \etal 2010 \PRA \textbf{81}
043831; Ou L and Kuang L 2007 \emph{Journ. Phys. B} \textbf{40} 1833
\bibitem{Chen} Chen X H \etal 2010 \OL \textbf{35}, 1166
\bibitem{boy}  Zerom P \etal 2011 \PRA \textbf{84} 061804
\bibitem{ipsc}  Puddu E \etal 2007 \OE  \textbf{32} 1132

\bibitem{gh2}   Pittman T \etal 1995 \PRA \textbf{52} R3429
\bibitem{av} Avella A 2015 \emph{J. Adv. Phys.} \textbf{4} 252

\bibitem{g} Gatti A, Brambilla E, and Lugiato L A 2003 \PRL \textbf{90} 133603; Gatti A, Brambilla E, and Lugiato L A 2004 \PRL \textbf{93} 093602; Gatti A, Brambilla E, and Lugiato L A 2004 \PRA \textbf{70} 013802; Cai Y and Zhu S 2004 \emph{Opt. Lett. }\textbf{29}  2716

\bibitem{gh31} Bennink R \etal 2002 \PRL \textbf{89}  113601
\bibitem{gh32} Ferri F \etal 2005 \PRL \textbf{94}  183602
\bibitem{gh33} Valencia A 2005 \PRL \etal \textbf{94} 063601
\bibitem{gh34} Zhai Y H \etal 2005 \PRA \textbf{72}  043805
\bibitem{gh35} Meyers R \etal 2008 \PRA \textbf{77} 041801
\bibitem{gh36} Chen X \etal 2010 \OL \textbf{35 }1166

\bibitem{sun} Liu X \etal 2014 \OL \textbf{39} 2314;   Karmakar S \etal 2011 - \emph{Proc. of CLEO Laser Applications to Photonic Applications, OSA Technical Digest (CD)} (Optical Society of America, 2011), paper QFD3
\bibitem{ghga} Ragy S and Adesso G 2012 \emph{Scientific Rep.} \textbf{2} 1

\bibitem{cgi} Shapiro J 2008 \PRA \textbf{78} 061802
\bibitem{cgin} Zhang E \etal 2015 \emph{J. Opt.} \textbf{17} 085602; Zafari M \etal  2014 \emph{J. Opt. }\textbf{16} 105405; Chen W and Chen X 2015 \emph{EPL} \textbf{109} 14001; Hardy D and Shapiro S 2013 \PRA \textbf{87} 023820


\bibitem{shsc} Shapiro J H, Venkatraman D and Wong F N C 2013 \emph{Scientific. Rep.} \textbf{3 }  1849


\bibitem{rM} Meyers R, Deacon K S and Shih Y 2008 \PRA \textbf{77}  041801.



\bibitem{tur} Gong W. and Han S.2011 \OL \textbf{36} 394; Bina M \etal 2013 \PRL \textbf{110} 083901; Dixon P. \etal 2011  \PRA \textbf{83} 051803; Meyers R \etal 2012 \emph{Applied Physics Letters} \textbf{100} 061126; Meyers R and Deacon K 2015 \emph{ Entropy} \textbf{17} 1508; Meyers R \etal 2012,\emph{ Appl. Phys. Lett.} \textbf{100} 061126; Meyers R \etal 2010,\emph{ Appl. Phys. Lett.} \textbf{98} 111115; Dixon P B \etal 2011 \PRA \textbf{83} 051803

\bibitem{al}  Meda A \etal 2015 \emph{Appl. Phys. Lett. }\textbf{106} 262405

\bibitem{gil1} Dong S \etal,  arXiv:1508.05248
\bibitem{gil2} Gong W \etal,  arXiv:1301.5767; Xu Y \etal 2015 \emph{Chin. Phys. B }\textbf{24} 124202
\bibitem{gil3} MorrisP \etal, \emph{ Nat. Comm.} DOI: 10.1038/ncomms6913
\bibitem{gil4} Aspden R \etal 2013 \emph{New Journal of Physics }\textbf{15} 073032

\bibitem{yogi} Cho Y 2012 \etal \OE \textbf{20} 5809

\bibitem{lug} Brambilla E \etal. 2008 \PRA \textbf{77} 053807

\bibitem{cal} Meda A \etal 2014 \emph{Appl. Phys. Lett.} \textbf{105} 101113; Brida G \etal 2006 \emph{J.Opt.Soc. Am B} \textbf{23} 2185; Brida G \etal 2008 \OE \textbf{16} 12550;
Brida G \etal 2010 \OE \textbf{18} 20572; Lindenthal M and Kofler J 2006  \emph{App. Opt.} \textbf{45} 6059; Perina J \etal 2012 \OL \textbf{37} 2475; A. Avella \etal in press.

\bibitem{s1} Nabors,C.D., and Shelby, R.M., 1990 \PRA 42 556
\bibitem{s2} Tapster,P.R.,  Seward, S.F., and  Rarity, J.G. 1991 \PRA 44 3266
\bibitem{s3} Souto Ribeiro, P.H., Schwob, C., Maitre, A. and Fabre, C. 1997 \OL 22 1893
\bibitem{s4} Bondani, M., Allevi, A., Zambra, G., Paris, M. and Andreoni A 2007 \PRA 76 013833; T.Ishkakov, M. Chekhova,
G.Leuchs 2009 \PRL 102, 183602
\bibitem{s5} Jedrkievicz, O. \etal. 2004 \PRL 93 243601
\bibitem{s6P} Pe\v{r}ina, J \etal 2012 \PRA { \bf 85}, 023816
\bibitem{s6} Blanchet J. \etal 2008 \PRL \textbf{101} 233604
\bibitem{s7} Ishkakov T \etal, arXiv:1511.08460.

\bibitem{ssn} Brida,G., Caspani, L., Gatti A., Genovese,M., Meda, A., and Ruo Berchera, I. 2009 \PRL 102, 213602
\bibitem{ssnexp} Brida G, Genovese M and Ruo Berchera I 2010 \emph{Nature Phot. } \textbf{4} 227
\bibitem{ssnpra} Brida G, Genovese M, Meda A and Ruo Berchera I 2011 \PRA \textbf{83}, 033811
\bibitem{bram}  Brambilla E, Gatti A, Bache M and Lugiato L A 2004\PRA \textbf{69}, 023802

\bibitem{spie} Ruo Bechera I \etal; Proc. SPIE 9225-1, Quantum Communications and Quantum Imaging XI, (2014)
\bibitem{rb16} Ruo Berchera I \etal in press

\bibitem{rqi}  Lawrie B and Pooser R 2013 \OE \textbf{21}7549

\bibitem{on13} Ono T \etal 2013 \emph{Nat. Comm.} \textbf{4} 2426

\bibitem{llo:08}  Lloyd S 2001 \emph{Science} {\bf 321}, 1463
\bibitem{da} D’Ariano G.,Lo Presti P. 2001 \PRL \textbf{86}  4195
\bibitem{tan:08}  Tan S {\it et al.} 2008 \PRL {\bf 101}, 253601
\bibitem{sha:09} Shapiro J H and Lloyd S 2009 \emph{New Journ. of Phys.} {\bf 11}, 063045
\bibitem{gua:09} Guha S and Erkmen B I 2009 \PRA {\bf 80}, 052310
\bibitem{ms} Sacchi M 2005 \PRA {\bf 71} 062340; Sacchi M 2005 \PRA {\bf 72}, 014305
\bibitem{l} Lopaeva E, Ruo Berchera I, Degiovanni I , Olivares S, Brida G, Genovese M 2013 \PRL \textbf{110} 153603
\bibitem{l1} Lopaeva E \etal 2014 \emph{Phys. Scr. }\textbf{\emph{T160}}  014026.

\bibitem{Ar} Arecchi T F 1965 \PRL 15, 912

\bibitem{ad} Ragy S \etal 2014 \emph{JOSA} \textbf{B 31} 2045


\bibitem{qiluut}  Zhang S \etal \PRA \textbf{89} 062309 (2014); Kang L \etal 2014 \emph{Chinese Physics Letters} \textbf{31} 034202; Lanzacorta M, Proc. SPIE 9461, Radar Sensor Technology XIX; and Active and Passive Signatures VI, 946113 (May 21, 2015); doi:10.1117/12.2177577; Zhang Sheng-Li \etal  2015 \emph{Chinese Physics Letters} \textbf{32}  090301; ShengLi Zhang \etal 2014
\PRA \textbf{90} 052308
\bibitem{mB}  Barzanjeh S \etal 2015 \PRL \textbf{114} 080503


\bibitem{nob} http://www.nobelprize.org/nobel-prizes/chemistry/laureates/2014/advanced-chemistryprize2014.pdf
\bibitem{po}Giovannetti V \etal 2009 \PRA \textbf{79} 013827
\bibitem{qm} Braun D and Popescu S 2014 \emph{Quantum Metrology} \textbf{2} 44

\bibitem{bip} Jacobson J \etal 1995 \PRL \textbf{74} 4835; Fonseca J \etal 1999 \PRL  \textbf{\emph{82}} 2868;
Edamatsu K, Shimizu R and Itoh T 2002 \PRL  \textbf{89} 213601 ; Santos I \etal 2003 \PRA \textbf{67} 033812;
\bibitem{noon}  Kim Y. \etal 2011 \OE 19 24957
\bibitem{lit}
Boto A \etal 2000 \PRL  \textbf{\emph{85}} 2733;
 D’Angelo M, Chekhova M and Shih Y 2001 \PRL  \textbf{\emph{87}} 013602;
Bennink R \etal 2004 \PRL  \textbf{\emph{92}} 033601; Demartini F \etal, \PRA \textbf{77} 012324; Tsang M. 2009 \PRL \textbf{102} 253601
\bibitem{xu} Xu D \etal 2015 \emph{Appl Phys Lett} \textbf{106} 171104
\bibitem{nw} Tsang M 2007 \PRA \textbf{75} 043813; Steuernagel O 2004 \emph{J.Opt. B} \textbf{6} S606
\bibitem{t} Tsang M 2009 \PRL \textbf{102} 253601
\bibitem{b} Shih H \etal 2011 \PRL \textbf{ 107} 083603
\bibitem{r} Rozema L \etal 2014 \PRL \textbf{112} 223602


\bibitem{post}Pregnell K and  Pegg D 2004 \emph{J. Mod. Opt.} \textbf{51} 1613; Hemmer P \etal 2006 \emph{Phys. Rev. Lett.} \textbf{96} 163603; Muthukrishnan \etal 2004 \emph{J.Opt B  }\textbf{ 6} S575;
Wang K \etal 2004 \PRA A \textbf{70} 041801; Cao D \etal 2015 \PRA \textbf{91} 053853 ;  Bjork G \etal \PRL 2001 \textbf{86} 4516; Nagasako E \etal 2001 \PRA \textbf{64} 043802 ; Shimizu R \etal 2003 \PRA \textbf{67} 041805

\bibitem{POST1}Bentley S and Boyd R 2004 \emph{Optics Express} \textbf{12} 5735 ;
\bibitem{post2} Guerrieri F \etal 2010 \PRL \textbf{105} 163602 .
\bibitem{d} Haroche S and Hartmann F, 1972 \PRA \textbf{6} 1280; Kyrola E and Stenholm S, 1977 \emph{Opt. Commun.} \textbf{22 }123;  Berman P R and Ziegler J 1977 Phys. Rev. A 15, 2042
(1977); Reid J and  Oka T 1977 \PRL \textbf{38} 67; Freund S M \etal 1975 \PRL \textbf{35}
1497; Bigelow N P and Prentiss M G 1990 \PRL \textbf{65} 555;  Tollett J J \etal 1990 \PRL \textbf{65}
559.
\bibitem{o} Joshi A and Osman K 2015 \emph{Eur. Phys. J. D} \textbf{69} 267; Chang H \etal 2006 \emph{Jour. Mod. Opt.} \textbf{53} 2271;
\bibitem{post3} Pe'er \etal 2004 \emph{Optics Express} \textbf{12} 6600
\bibitem{yo} Oh J \etal 2013 \OL \textbf{ 38}   682
\bibitem{post4} Mouradian S \etal 2011 \OE \textbf{19} 5480
\bibitem{wu} Zhai Y \etal 2005 \PRA  \textbf{72} 043805
\bibitem{th} Zhai Y \etal 2014 \emph{Appl. Phys. Lett.} \textbf{105} 101104; Xu-Ri Y \etal 2014 \emph{CHINESE PHYSICS B}  \textbf{24}  044203; Bobrov I \etal 2014 \PRA \PRA \textbf{89} 043814;

\bibitem{sc} Liu R \etal 2014 \emph{Scientific. Rep.} \textbf{4 } 1

\bibitem{zan07} Thiel C \etal 2007 \PRL \textbf{99} 133603
\bibitem{zan09} Thiel C \etal 2009 \PRA \textbf{80} 013820
\bibitem{zan12} Oppel S \etal 2012 \PRL \textbf{109} 233603

\bibitem{or} Schwartz O and Oron D 2012 \PRA \textbf{85 } 033812; Schwartz O \etal 2013 \emph{Nano Lett.} \textbf{13} 5832 


\bibitem{pt} Gatto Monticone D \etal 2014 \PRL \textbf{113} 143602

\bibitem{v2} Steinert S \etal 2013 \emph{Nature Comm.} \textbf{4} 1607; Hall L T \etal 2012 \emph{Scientific Rep. } \textbf{2} 401; V. Carabelli \etal in press.
\bibitem{sr} Forneris J 2015 \emph{Scientific Reports} \textbf{5} 15901 and ref.s therein.

\bibitem{kf}Kolobov M and Fabre C 2000 \PRL \textbf{85 } 3789
\bibitem{ta13} Taylor M \etal 2013 \emph{Nat. Phot} \textbf{7} 229
\bibitem{ta14} Taylor M \etal 2014 \PRX \textbf{4} 011017

\bibitem{bo08} Boyer v \etal 2008 \emph{Science} \textbf{321} 544
\bibitem{mo05} Mosset A \etal 2005 \PRL \textbf{94} 223603
\bibitem{le14} Lemos G \etal 2014 \emph{Nature} \textbf{512} 409

\bibitem{alf}Cialdi, S \etal 2011 \PRA \textbf{84} 043817; Jachura M and Chrapkiewicz R 2015 \OL \textbf{40}  1540; Lopez L \etal 2008 \PRL \textbf{100} 013604; Delaubert V \etal 2008  \emph{EPL}  \textbf{81} 44001; P\`erez-Delgado C \etal   2012 \PRL \textbf{109} 123601; Treps N \etal 2002 \PRL \textbf{88} 203601;  Santos I F, Aguirre-Gomez J G and Padua S 2008 \PRA  \textbf{77} 043832; Sokolov I V \etal (2001) \emph{Optic. Comm} \emph{193} 195; Abouraddy A \etal 2004 \PRL \textbf{93} 213903; Abouraddy A \etal 2002  \emph{Journ. of Am opt. Soc. B}\textbf{19} 1174; Andreoni A \etal 2006
\PRA \textbf{74} 053802 




\end{thebibliography}
\end{document}